\documentclass[11pt]{article}

\usepackage[utf8]{inputenc}
\usepackage{amsmath}
\usepackage{amssymb}
\usepackage{amsthm}
\usepackage{color,soul}
\usepackage{tikz}
\usetikzlibrary{topaths,calc}
\usepackage{subfigure}
\usepackage{cite}

\def\QED{\mbox{\rule[0pt]{1.5ex}{1.5ex}}}
\def\proof{\noindent\hspace{2em}{\it Proof: }}

\def\QED{\mbox{\rule[0pt]{1.5ex}{1.5ex}}}

\newtheorem{theorem}{Theorem}

\newtheorem{lemma}{Lemma}

\newcommand\blfootnote[1]{%
  \begingroup
  \renewcommand\thefootnote{}\footnote{#1}%
  \addtocounter{footnote}{-1}%
  \endgroup
}

\title{Secure Summation: Capacity Region, \\
Groupwise Key, and Feasibility}
\author{Yizhou Zhao and Hua Sun}
\date{}

\begin{document}
\maketitle

\blfootnote{
Yizhou Zhao (email: yizhouzhao@my.unt.edu) and Hua Sun (email: hua.sun@unt.edu) are with the Department of Electrical Engineering at the University of North Texas.}

\begin{abstract}
The secure summation problem is considered, where $K$ users, each holds an input, wish to compute the sum of their inputs at a server securely, i.e., without revealing any information beyond the sum even if the server may collude with any set of up to $T$ users. First, we prove a folklore result for secure summation - to compute $1$ bit of the sum securely, each user needs to send at least $1$ bit to the server, each user needs to hold a key of at least $1$ bit, and all users need to hold collectively some key variables of at least $K-1$ bits. Next, we focus on the symmetric groupwise key setting, where every group of $G$ users share an independent key. We show that for symmetric groupwise keys with group size $G$, when $G > K-T$, the secure summation problem is not feasible; when $G \leq K-T$, to compute $1$ bit of the sum securely, each user needs to send at least $1$ bit to the server and the size of each groupwise key is at least $(K-T-1)/\binom{K-T}{G}$ bits. Finally, we relax the symmetry assumption on the groupwise keys and the colluding user sets; we allow any arbitrary group of users to share an independent key and any arbitrary group of users to collude with the server. For such a general groupwise key and colluding user setting, we show that secure summation is feasible if and only if the hypergraph, where each node is a user and each edge is a group of users sharing the same key, is connected after removing the nodes corresponding to any colluding set of users and their incident edges. 
\end{abstract}

\newpage
\allowdisplaybreaks
\section{Introduction}
The advent of the modern information age is enabled by pervasive networked communication and computation devices, which accelerate data exchange at an unprecedented pace and bring security concerns to the forefront. The need to securely perform distributed computation tasks has thus increased tremendously. This work is particularly motivated by the secure aggregation problem \cite{aggregation, aggregation_log, aggregation_turbo, aggregation_fast, Zhao_Sun_Aggregate, aggregation_light, aggregation_swift, WSJC_Groupwise, aggregation_rosnes, aggregation_jun}, which arises recently in federated learning and the core is to securely compute the sum of the inputs available at a number of users without revealing any additional information to a server. While secure aggregation is usually involved with more practical elements that are crucial for machine learning applications, such as user dropouts, peer-to-peer communication among the users etc., in this work we focus on an elemental information theoretic model that is possibly the simplest while capturing the core of secure sum computation (referred to as secure summation), and wish to understand its fundamental limits on communication and randomness cost.

\begin{figure}[h]
\centering
\begin{tikzpicture}
    \node (u1) at (0,4) {};
    \node (u2) at (0,2.5) {};
    \node at (0.5,1.9) {$\vdots$};
    \node (uK) at (0,0) {};
    \node (server) at (8,2.25) {};
    \node (server1) at ($(server)+(0.1,-0.5)$) {};
    \node (server2) at ($(server)+(0.1,-0.2)$) {};
    \node (server3) at ($(server)+(0.1,0.1)$) {};
    \node (server4) at ($(server)+(0.1,0.4)$) {};
    \filldraw ($(u1)$)
    to[out=90,in=180] ($(u1) + (0.5,0.5)$)
    to[out=0,in=90] ($(u1) + (1,0)$);
    \fill ($(u1) + (0.5,0.8)$) circle(0.3);
    \filldraw ($(u2)$)
    to[out=90,in=180] ($(u2) + (0.5,0.5)$)
    to[out=0,in=90] ($(u2) + (1,0)$);
    \fill ($(u2) + (0.5,0.8)$) circle(0.3);
    \filldraw ($(uK)$)
    to[out=90,in=180] ($(uK) + (0.5,0.5)$)
    to[out=0,in=90] ($(uK) + (1,0)$);
    \fill ($(uK) + (0.5,0.8)$) circle(0.3);
    \filldraw ($(server)+(0,-0.6)$) rectangle ($(server)+(1,0.7)$);
    \foreach \v in {1,2,...,4} {
        \filldraw [white] (server\v) rectangle ($(server\v)+(0.8,0.2)$);
        \filldraw ($(server\v)+(0.3,0.08)$) rectangle ($(server\v)+(0.75,0.12)$);
        \fill ($(server\v)+(0.15,0.1)$) circle (0.05);
    }
    \draw [rounded corners,->, line width=1pt,shorten >=10pt]($(u1) + (1.15,0.2)$) -- ($(u1) + (5.5,0.2)$)
    -- (server);
    \draw [rounded corners,->, line width=1pt,shorten >=10pt]($(u2) + (1.15,0.2)$) -- ($(u2) + (5.5,0.2)$)
    -- (server);
    \draw [rounded corners,->, line width=1pt,shorten >=10pt]($(uK) + (1.15,0.2)$) -- ($(uK) + (5.5,0.2)$)
    -- (server);
    \node at ($(u1) + (1,0.5)$) [right]{$X_1=W_1+N_1$};
    \node at ($(u2) + (1,0.5)$) [right]{$X_2=W_2+N_2$};
    \node at ($(uK) + (1,0.5)$) [right]{$X_K=W_K-\sum_{k=1}^{K-1} N_k$};
    \node at ($(u1) + (0.5,-0.2)$) {User $1$};
    \node at ($(u2) + (0.5,-0.2)$) {User $2$};
    \node at ($(uK) + (0.5,-0.2)$) {User $K$};
    \node at ($(u1) + (0,0.5)$) [left]{$W_1,Z_1$};
    \node at ($(u2) + (0,0.5)$) [left]{$W_2,Z_2$};
    \node at ($(uK) + (0,0.5)$) [left]{$W_K,Z_K$};
    \node at ($(server)+(0.5,-0.9)$) []{Server};
    \node at ($(server)+(2.5,0.25)$) []{only learn};
    \node at ($(server)+(1,-0.2)$) [right]{$W_1+W_2+\cdots+W_K$};
\end{tikzpicture}
\caption{The secure summation problem and an optimal protocol.}
\label{fig:model}
\end{figure}
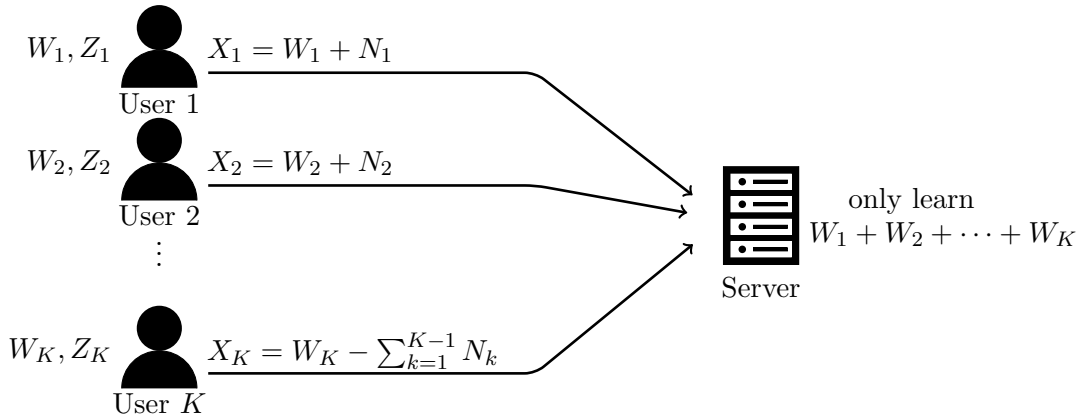

In the secure summation problem (see Fig.~\ref{fig:model}), we have $K$ users, each holds an independent input $W_k, k \in \{1,2,\cdots,K\}$ over some finite field and a key variable $Z_k$ that is independent of $W_k$ and serves to ensure security. Each user is connected to the server through an orthogonal noiseless link, over which User $k$ can send a message $X_k$. After receiving one message from each user (i.e., from $X_1, \cdots, X_K$), the server must be able to compute the sum of all $W_k$ but learn no extra information in the information theoretic sense even if the server may collude with at most\footnote{Without loss of generality, we assume $0 \leq T \leq K-2$ because when the server colludes with $K-1$ or $K$ users, the server can learn all inputs $W_k$ such that there is nothing to hide. As a result, when $T= K-1$ or $K$, the results are the same as those when $T=K-2$.} $T$ arbitrary users.

A simple protocol (see e.g., \cite{Hayashi_Koshiba, Zhao_Sun_Expand}) is depicted in Fig.~\ref{fig:model}, where the key variables are assigned such that their sum is zero, i.e., $Z_k = N_k, k \in \{1,\cdots, K-1\}, Z_K = -\sum_{k=1}^{K-1}Z_k$ and $N_k$ are i.i.d. and uniform over the same field as the inputs. Interestingly, this protocol turns out to be information theoretically optimal in terms of both the communication cost and the key size - in order to securely compute $1$ symbol of the sum $W_1+\cdots+W_K$, each user must communicate at least $1$ symbol of $X_k$ to the server, each user must hold a key $Z_k$ that is at least $1$ symbol, and all users must collectively hold key variables whose joint entropy $H(Z_1,\cdots,Z_K)$ is at least $K-1$ symbols. Note that the optimal communication and key rates do not depend on the maximum number of colluding users, $T$. While we expect this result to be known and indeed the proof follows relatively straightforwardly from existing work (e.g., \cite{Zhao_Sun_Aggregate, WSJC_Secure}), we do not find this result explicitly published anywhere and therefore give a full proof in this work (refer to Theorem \ref{thm:coded}), which also turns out to be useful in our next set of results.

In the basic model of secure summation, the keys $Z_k$ are allowed to be arbitrarily correlated and thus typically are assigned by a trusted third-party or generated through interactive communication among the users, which might be occasionally restrictive in practice. Observing that generating a key securely among a group of users is a well studied primitive \cite{Maurer_Key, Ahlswede_Csiszar_CR, Csiszar_Narayan, Gohari_Anantharam, Chan_Zheng}, we proceed to consider the symmetric groupwise key setting, where every $G$ users from the set $\mathcal{G}$ share an independent key $S_{\mathcal{G}}$ and all such keys have the same size. For example, suppose $K = 3, G = 2$. Then we have $3$ groupwise keys $S_{\{1,2\}}, S_{\{1,3\}}, S_{\{2,3\}}$ and the key variable at User 1 is  $Z_1 = (S_{\{1,2\}}, S_{\{1,3\}})$.

For the secure summation problem with symmetric groupwise keys, we fully characterize the optimal communication and key rates in Theorem \ref{thm:groupwise}. We have two regimes - when $G > K-T$, the secure summation problem is not information theoretically feasible, intuitively because too many keys are known to a colluding user set such that leakage cannot be avoided; when $G \leq K-T$, to securely compute $1$ symbol of the sum, the optimal communication rate remains the same as the arbitrarily coded key case, i.e., groupwise keys do not hurt and each user must communicate at least $1$ symbol to the server, and the minimum size of the symmetric groupwise key $S_{\mathcal{G}}$ is $(K-T-1)/\binom{K-T}{G}$ symbols. The achievable scheme is based on a randomized key construction and the converse builds upon that of the arbitrarily coded key setting considered in Theorem \ref{thm:coded} and incorporates the groupwise key constraint (refer to Theorem \ref{thm:groupwise}). 

\begin{figure}[h]
\centering
\subfigure[]
{
\centering
\begin{tikzpicture}
    \node (v1) at (0,2) {};
    \node (v2) at (2,2) {};
    \node (v3) at (0,0) {};
    \node (v4) at (2,0) {};
    \begin{scope}[fill opacity=0.6]
    \filldraw[fill=red!70] ($(v1)+(-0.5,0)$) 
        to[out=90,in=180] ($(v1) + (0.8,0.5)$) 
        to[out=0,in=145] ($(v2) + (0.5,0.5)$) 
        to[out=-35,in=0] ($(v4) + (0,-0.5)$)
        to[out=180,in=0] ($(v2) + (-0.5,0)$)
        to[out=180,in=270] ($(v1)+(-0.5,0)$);
    \filldraw[fill=green!80] ($(v3)+(0,-0.5)$) 
        to[out=180,in=270] ($(v3)+(-0.5,0)$) 
        to[out=90,in=180] ($(v2) + (0,0.5)$)
        to[out=0,in=40] ($(v2) + (0.5,-0.5)$)
        to[out=210,in=30] ($(v2) + (-1,-1)$)
        to[out=210,in=0] ($(v3) + (0,-0.5)$);
    \filldraw[fill=blue!50] ($(v3)+(-0.3,0.3)$) 
        to[out=90,in=180] ($(v3)+(1,0.2)$)
         to[out=0,in=90] ($(v4)+(0.5,0)$)
        to[out=270,in=270] ($(v3) + (-0.3,0.3)$);
    \end{scope}
    \fill (v1) circle (0.1) node [above] {$v_1$};
    \fill (v2) circle (0.1) node [right] {$v_2$};
    \fill (v3) circle (0.1) node [above] {$v_3$};
    \fill (v4) circle (0.1) node [right] {$v_4$};
    \node at ($(v4)+(0.5,1)$) {$e_1$};
    \node at (0.3,1) {$e_2$};
    \node at (1,-0.1) {$e_3$};
\end{tikzpicture}}
\subfigure[]
{
\centering
\begin{tikzpicture}
    \node (v1) at (0,2) {};
    \node (v2) at (2,2) {};
    \node (v3) at (0,0) {};
    \begin{scope}[fill opacity=0.6]
    \filldraw[fill=green!80] ($(v3)+(0,-0.5)$) 
        to[out=180,in=270] ($(v3)+(-0.5,0)$) 
        to[out=90,in=180] ($(v2) + (0,0.5)$)
        to[out=0,in=40] ($(v2) + (0.5,-0.5)$)
        to[out=210,in=30] ($(v2) + (-1,-1)$)
        to[out=210,in=0] ($(v3) + (0,-0.5)$);
    \filldraw[fill=blue!50,opacity=0] ($(v3)+(-0.3,0.3)$) 
        to[out=90,in=180] ($(v3)+(1,0.2)$)
         to[out=0,in=90] ($(v4)+(0.5,0)$)
        to[out=270,in=270] ($(v3) + (-0.3,0.3)$);
    \end{scope}
    \fill (v1) circle (0.1) node [above] {$v_1$};
    \fill (v2) circle (0.1) node [right] {$v_2$};
    \fill (v3) circle (0.1) node [above] {$v_3$};
    \node at (0.3,1) {$e_2$};
\end{tikzpicture}
}
\subfigure[]
{
\centering
\begin{tikzpicture}
    \node (v1) at (0,2) {};
    \node (v2) at (2,2) {};
    \node (v4) at (2,0) {};
    \begin{scope}[fill opacity=0.6]
    \filldraw[fill=red!70] ($(v1)+(-0.5,0)$) 
        to[out=90,in=180] ($(v1) + (0.8,0.5)$) 
        to[out=0,in=145] ($(v2) + (0.5,0.5)$) 
        to[out=-35,in=0] ($(v4) + (0,-0.5)$)
        to[out=180,in=0] ($(v2) + (-0.5,-0.5)$)
        to[out=180,in=270] ($(v1)+(-0.5,0)$);
    \filldraw[fill=blue!50,opacity=0] ($(v3)+(-0.3,0.3)$) 
        to[out=90,in=180] ($(v3)+(1,0.2)$)
         to[out=0,in=90] ($(v4)+(0.5,0)$)
        to[out=270,in=270] ($(v3) + (-0.3,0.3)$);
    \end{scope}
    \fill (v1) circle (0.1) node [above] {$v_1$};
    \fill (v2) circle (0.1) node [right] {$v_2$};
    \fill (v4) circle (0.1) node [right] {$v_4$};
    \node at ($(v4)+(0.5,1)$) {$e_1$};
\end{tikzpicture}
}
\caption{(a) The hypergraph representation of $3$ groupwise keys $S_{\{1,2,4\}}, S_{\{2,3\}}, S_{\{3,4\}}$. (b) The remaining hypergraph after removing $v_4$. 
(c) The remaining hypergraph after removing $v_3$.
}
\label{fig:hypergraph}
\end{figure}
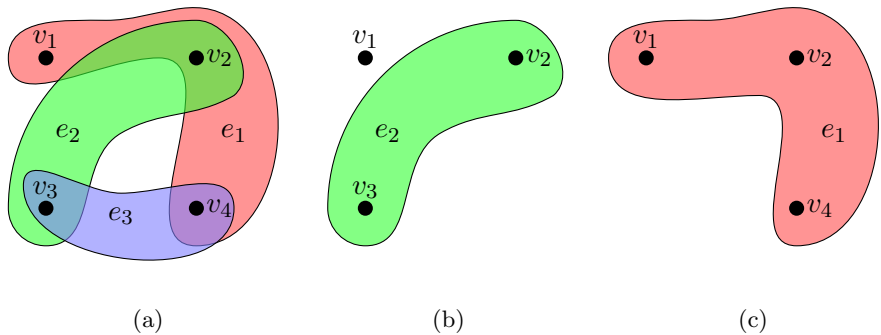

Finally, we relax the symmetry assumption on the groupwise keys (i.e., every $G$ users share a key) and on the colluding user sets (i.e., every $T$ users might be colluding). For any group of users, we allow them to share an independent key and/or to be a colluding user set. We wish to understand for which groupwise key and colluding user pattern, the secure summation problem is feasible. The necessary and sufficient condition is obtained in Theorem \ref{thm:feasible}, which is stated in terms of a hypergraph representation of the key pattern, explained now. For example, suppose we have $K=4$ users and $3$ groupwise keys $S_{\{1,2,4\}}$ (i.e., a key shared by User 1, 2, 4), $S_{\{2,3\}}, S_{\{3,4\}}$. Representing this groupwise key setting with a hypergraph (see Fig.~\ref{fig:hypergraph}(a)), we have $4$ nodes $v_1, v_2, v_3, v_4$, where $v_k$ corresponds to User $k$, and $3$ hyperedges $e_1, e_2, e_3$, each corresponds to a set of users sharing a same key (e.g., $e_1$ corresponds to $S_{\{1,2,4\}}$ such that $e_1$ is incident with $v_1, v_2, v_4$).

Equipped with the hypergraph key representation, we are ready to state the result. We show that the secure summation problem is feasible if and only if after removing the node set that corresponds to any colluding set of users and their incident edges, the remaining hypergraph is connected. Consider again the example in Fig.~\ref{fig:hypergraph}(a). Suppose User 4 might collude with the server. To see if this secure summation problem is feasible, we remove $v_4$ and its incident edges from the hypergraph and obtain Fig.~\ref{fig:hypergraph}(b), which is not connected as $v_1$ is isolated, so we conclude that secure summation is not feasible. The intuition is that there are too few keys that are hidden from the colluding user set so that it is not possible to fully protect the desired sum. As another example, suppose User 3 might be colluding, removing which we have a connected subgraph Fig.~\ref{fig:hypergraph}(c) so that secure summation is feasible. Here we have sufficient keys to ensure both security and decodability of the sum. When there are multiple possible colluding user sets, for a secure summation problem to be feasible, we need to ensure the connectivity of the hypergraph after removing each colluding user set.

\bigskip
{\em Notation: For positive integers $K_1$, $K_2$, $K_1 < K_2$, we use the notation $[K_1:K_2] \triangleq \{K_1, K_1+1,\cdots, K_2\}$ and $[1:K_2]$ is abbreviated as $[K_2]$. 
The notation $|\mathcal{A}|$ is used to denote the cardinality of a set $\mathcal{A}$. For two sets $\mathcal{A}$ and $\mathcal{B}$, we use $\mathcal{A}\backslash\mathcal{B}$ to denote the set of elements that belong to $\mathcal{A}$ but not $\mathcal{B}$. The notation $\binom{\mathcal{A}}{G}$ is used to denote all subsets of $\mathcal{A}$ with cardinality $G$, i.e., $\binom{\mathcal{A}}{G} \triangleq \{\mathcal{G}:\mathcal{G}\subset{\mathcal{A}},|\mathcal{G}|=G\}$. 
}

\section{Problem Statement}\label{section:model}
The secure summation problem involves one server and $K$ users, where $K \geq 2$ and User $k \in [K]$ holds an input vector $W_k$ and a key variable $Z_k$. The input vectors $\left(W_k\right)_{k\in[K]}$ are independent. Each $W_k$ is an $L \times 1$ column vector and the $L$ elements are i.i.d. uniform symbols from the finite field $\mathbb{F}_q$. $\left(W_k\right)_{k\in[K]}$ is independent of $\left(Z_k\right)_{k\in[K]}$. 
\begin{eqnarray}
   && H\left(\left(W_k\right)_{k\in[K]},
    \left(Z_k\right)_{k\in[K]}\right)=
    \sum_{k\in[K]} H\left(W_k \right) +
    H\left(\left(Z_k\right)_{k\in[K]} \right), \label{ind} \\
   && H(W_k) = L ~(\mbox{in $q$-ary units}), ~\forall k \in [K]. \label{h2}
\end{eqnarray}
Each $Z_k$ is comprised of $L_Z$ symbols from $\mathbb{F}_q$. The key variables can be arbitrarily correlated and are a function of a source key variable $Z_\Sigma$, which is comprised of $L_{Z_{\Sigma}}$ symbols from $\mathbb{F}_q$.
\begin{eqnarray}
	H\left(\left(Z_k\right)_{k\in[K]} \Big| Z_\Sigma\right) = 0.\label{total rand}
\end{eqnarray}
The communication protocol includes one message from each user to the server. Specifically, User $k$ sends a message $X_k$, $k \in [K]$ to the server. The message $X_k$ is a function of $W_k, Z_k$ and consists of $L_X$ symbols from $\mathbb{F}_q$.
\begin{eqnarray}
    H\left(X_k | W_k, Z_k\right) = 0, \forall k \in [K].\label{message}
\end{eqnarray}
From all messages, the server must be able to recover the desired sum $\sum_{k \in [K]} W_k$ with no error.
\begin{eqnarray}
    \mbox{[Correctness]}~~~H\left(\sum_{k\in[K]} W_k \Bigg| \left(X_k\right)_{k\in[K]} \right) = 0.\label{corr}
\end{eqnarray}
We impose that security must be guaranteed even if the server may collude with any set of at most $T$ users, where $0 \leq T \leq K-2$.
Specifically, security refers to the constraint that the server cannot infer any additional information about $\left(W_k\right)_{k\in[K]}$ beyond that contained in the desired sum and known from the colluding users. That is, the following security constraint must be satisfied for any $\mathcal{T}$, where $\mathcal{T} \subset{[K]}, |\mathcal{T}|\leq T$.
\begin{eqnarray}
    \mbox{[Security]}~~~I\left(\left(W_k\right)_{k\in[K]}; \left(X_k\right)_{k\in[K]} \Bigg| \sum_{k\in [K]} W_k, \left( W_k, Z_k \right)_{k\in\mathcal{T}} \right) = 0.\label{security}
\end{eqnarray}
The communication {\em rate} $R$ characterizes how many symbols each message contains per input symbol, and is defined as follows.
\begin{eqnarray}
	R \triangleq \frac{L_X}{L}. \label{rate:R}
\end{eqnarray}
The individual (total) key {\em rate} $R_{Z}$ ($R_{Z_{\Sigma}}$) characterizes how many symbols each key variable (the source key variable) contains per input symbol, and is defined as follows.
\begin{eqnarray}
    R_{Z} \triangleq \frac {L_{Z}}{L}, ~R_{Z_{\Sigma}} \triangleq \frac {L_{Z_{\Sigma}}}{L}. \label{rate:R_Z}
\end{eqnarray}
The rate tuple $(R,R_{Z},R_{Z_{\Sigma}})$ is said to be achievable if there exists a secure summation scheme, for which the correctness constraint (\ref{corr}) and the security constraint (\ref{security}) are satisfied, and the communication rate, the individual key rate, and the total key rate are no greater than $R, R_{Z}$, and $R_{Z_{\Sigma}}$, respectively. The closure of the set of all achievable rate tuples is called the optimal rate region (i.e., capacity region), denoted as $\mathcal{R}^{*}$.

\subsection{Symmetric Groupwise Keys}\label{subsection:model2}
The symmetric groupwise key setting refers to a specific type of joint distribution of the keys, where every $1\leq G\leq K$ users share an equal-size independent key.
Consider $\binom{K}{G}$ independent random variables $S_{\mathcal{G}}, \mathcal{G} \in \binom{[K]}{G}$ and each $S_{\mathcal{G}}$ is comprised of $L_S$ i.i.d. uniform symbols from $\mathbb{F}_q$. 
\begin{eqnarray}
    H\left(\left(S_{\mathcal{G}}\right)_{\mathcal{G} \in \binom{[K]}{G}}\right)
    = \sum_{\mathcal{G} \in \binom{[K]}{G}} H(S_{\mathcal{G}}) = \binom{K}{G} L_S. \label{ind_S}
\end{eqnarray}
For example, when $K=4, G=3$, we have $S_{\{1,2,3\}},S_{\{1,2,4\}},S_{\{1,3,4\}},S_{\{2,3,4\}}$. The key variable $S_{\mathcal{G}}$ is shared by users in $\mathcal{G}$ so that the key variable held by User $k$, $Z_k$ is given by
\begin{eqnarray}
	Z_k=\left(S_{\mathcal{G}}\right)_{k \in \mathcal{G}, \mathcal{G} \in \binom{[K]}{G}}, \forall k \in [K]. \label{Z_individual}
\end{eqnarray}
For example, when $K=4, G=3$, $Z_3=(S_{\{1,2,3\}},S_{\{1,3,4\}},S_{\{2,3,4\}})$. 
The groupwise key rate $R_S$ characterizes how many symbols each groupwise key variable contains per input symbol, and is defined as follows.
\begin{eqnarray}
    R_S \triangleq \frac{L_S}{L}. \label{rate:R_S}
\end{eqnarray}
Note that due to the symmetry assumption on the groupwise keys, the individual and total key rates $R_Z, R_{Z_{\Sigma}}$ can be readily obtained from $R_S$, i.e., $R_{Z} = \binom{K-1}{G-1} R_{S}, R_{Z_{\Sigma}} = \binom{K}{G} R_{S}$. Thus for the symmetric groupwise key setting, the rate tuple $(R,R_{Z},R_{Z_{\Sigma}})$ can be more succinctly captured by $(R,R_{S})$, which will be adopted. The closure of the set of all achievable rate tuples $(R,R_{S})$ is called the optimal rate region, denoted as $\mathcal{R}^*_{g}$.

\subsection{General Groupwise Keys and Colluding Users}\label{subsection:graph}
Generalizing the symmetric groupwise key setting to include any possilbe subsets of users, we have the general groupwise key setting, i.e., any subset $\mathcal{G} \subset [K]$ of users may share an independent key variable $S_{\mathcal{G}}$ of size $L_S$ symbols\footnote{We are interested in the feasibility of the secure summation problem for the general groupwise key setting so that we may assume that each key has the same size with no loss.} from $\mathbb{F}_q$. Denote the family (set) of all sets of users that share a key by $\overline{\mathcal{G}}$. Then we can represent such groupwise keys by a hypergraph $H = (\mathcal{V}, \mathcal{E})$, defined as follows. The node set $\mathcal{V} = \{v_1, \cdots, v_K\}$ contains $K$ nodes, where $v_k, k \in [K]$ represents User $k$; the edge set $\mathcal{E} = \{e_1, \cdots, e_{\left|\overline{\mathcal{G}}\right|}\}$ contains $\left|\overline{\mathcal{G}}\right|$ (hyper)edges, where each edge $e_i, i \in [\left|\overline{\mathcal{G}}\right|]$ represents a group of users that share a same key, i.e., $e_i = \{(v_j)_{j \in \mathcal{G}_i}\}$ where $\mathcal{G}_i$ is the $i$-th element of $\overline{\mathcal{G}}$ (we may take any order of the elements in the set). 
A key hypergraph example can be found in Fig.~\ref{fig:hypergraph}(a), where $\overline{\mathcal{G}} = \{\{1,2,4\}, \{2,3\}, \{3,4\}\}$ and $e_1=\{v_1,v_2,v_4\}, e_2 = \{v_2, v_3\}, e_3 = \{v_3, v_4\}$. A hypergraph $H = (\mathcal{V}, \mathcal{E})$ (with at least two nodes) is said to be {\em connected} if for any $\mathcal{V}_1 \subset \mathcal{V}, 0< |\mathcal{V}_1| < |\mathcal{V}|$, there exists one edge $e \in \mathcal{E}$ such that $e \not\subset
 \mathcal{V}_1$ and $e \not\subset \mathcal{V}\backslash\mathcal{V}_1$, i.e., for any two-part partition of the node set, there exists at least one edge that connects the two parts.

Similarly, we generalize the symmetric colluding user setting to include any possible subsets of users, i.e., any subset $\mathcal{T} \subset [K]$ of users may collude with the server. Denote the family of all colluding user set by $\overline{\mathcal{T}}$ and the security constraint is the same as that in (\ref{security}), but it holds for all $\mathcal{T} \in \overline{\mathcal{T}}$. We will study the feasibility of secure summation, i.e., if the optimal rate region\footnote{Some explanation on the notation - $\mathcal{R}^*$ denotes the optimal rate region with arbitrarily coded keys, $\mathcal{R}_g^*$ denotes the optimal rate region with symmetric groupwise keys, and $\mathcal{R}_{gc}^*$ denotes the optimal rate region with general groupwise keys and colluding user sets.} $\mathcal{R}_{gc}^*$ (closure of achievable rate tuples $(R, R_S)$) is empty or not. The condition turns out to be related to the connectivity of the key hypergraph $H$ after removing a colluding user set and the known keys, which is denoted by $H[\mathcal{V}\backslash \{(v_k)_{k \in \mathcal{T}}\}]$, i.e., the induced subgraph of $H$ with node set $\mathcal{V}\backslash \{(v_k)_{k \in \mathcal{T}}\}$.

\section{Results}
In this section, we summarize our main results along with key observations. 

\subsection{Secure Summation: Capacity Region}
The optimal rate region of secure summation is characterized in Theorem \ref{thm:coded}, presented below.
\begin{theorem}\label{thm:coded}
For the secure summation problem with $K \geq 2$ users and at most $0\leq T \leq K-2$ colluding users, the optimal rate region is
\begin{eqnarray}
    \mathcal{R}^* =
    \left\{ \left(R, R_{Z}, R_{Z_{\Sigma}}\right) : R \geq 1, R_{Z} \geq 1, R_{Z_{\Sigma}} \geq K-1\right\}.
\end{eqnarray}
\end{theorem}

An intuitive explanation of Theorem \ref{thm:coded} may be seen as follows. To securely compute $1$ symbol of the sum at the server, each user needs to send at least $1$ symbol (which carries its own input), each user needs to hold a key of at least $1$ symbol (to protect the $1$ symbol transmit message), and all users must hold some key variables of at least $K-1$ symbols (note that the server must be able to only decode the $1$ symbol sum from all $K$ symbols received, so the remaining $K-1$ symbols must be fully protected by some key variables). The proof of Theorem \ref{thm:coded} is presented in Section \ref{pf:coded}. 

\subsection{Secure Summation with Symmetric Groupwise Keys: Capacity Region}

The optimal rate region of secure summation with symmetric groupwise keys is characterized in Theorem \ref{thm:groupwise}, presented below.

\begin{theorem}\label{thm:groupwise}
For the secure summation problem with $K \geq 2$ users, at most $0 \leq T \leq K-2$ colluding users, and symmetric groupwise keys of group size $G$, the optimal rate region is
\begin{eqnarray}
    \mathcal{R}_g^* = \left\{
    \begin{array}{cl}
        \emptyset & ~\mbox{when}~G > K-T, \\
        \left\{ \left(R, R_{S}\right) : R \geq 1, R_{S} \geq \cfrac{K-T-1}{\binom{K-T}{G}} \right\}& ~\mbox{when}~ G \leq K-T.
    \end{array}
    \right.
\end{eqnarray}
\end{theorem}

An intuitive explanation of Theorem \ref{thm:groupwise} may be seen as follows. When $G > K-T$, consider any set of $T$ colluding users $\mathcal{T}$, who know all the keys as any groupwise key $S_{\mathcal{G}}, |\mathcal{G}| = G$ involves at least one user in $\mathcal{T}$. As all the keys are known to the server (through user collusion), nothing can be hidden from the server, which violates the security constraint so that secure summation is not feasible. When $G \leq K-T$, suppose for now we have deleted $T$ colluding users and the keys known to them, leaving us with a secure summation problem with $K-T$ users and symmetric groupwise keys of size $G$ among them. For this problem with $K-T$ users, from Theorem \ref{thm:coded}, the total key size should be at least $K-T-1$, i.e., $K-T-1 \leq \binom{K-T}{G} R_S$ and we have the desired converse on the key rate. Guided by the converse, we will design a vector linear scheme, where the keys are precoded by matrices with proper size such that the sum of the keys from a given group is equal to zero (to guarantee correctness) and the overall key blocks are sufficiently generic (i.e., full rank, to guarantee security). The details are deferred to the proof of Theorem \ref{thm:groupwise}, presented in Section \ref{sec:groupwise}.

\subsection{Secure Summation with General Groupwise Keys and Colluding Users: Feasibility Condition}

The feasibility condition of secure summation with general groupwise keys and colluding users is characterized in Theorem \ref{thm:feasible}, presented below.

\begin{theorem}\label{thm:feasible}
For the secure summation problem with groupwise key hypergraph $H=(\mathcal{V},\mathcal{E})$ and colluding user set family $\overline{\mathcal{T}}$, 
\begin{eqnarray}
\mathcal{R}_{gc}^* \neq \emptyset ~\mbox{if and only if}~H[\mathcal{V}\backslash\{\left(v_k\right)_{k \in\mathcal{T}}\}]~\mbox{is connected for any}~\mathcal{T} \in \overline{\mathcal{T}}.     
\end{eqnarray}
\end{theorem}

An intuitive explanation of Theorem \ref{thm:feasible} may be seen as follows. On the one hand, when the induced subgraph of $H$ is not connected after removing some colluding user set, the remaining keys are not sufficiently correlated to protect the desired sum as there exist two sets of users (a two-part partition of the subgraph) whose keys are independent (note that there is no edge connecting the two parts, i.e., no key is known to both parts). Note that the $G > K-T$ case of Theorem \ref{thm:groupwise} is covered by Theorem \ref{thm:feasible} as the edge set of $H[\mathcal{V}\backslash\{\left(v_k\right)_{k \in\mathcal{T}}\}]$ with symmetric groupwise keys is empty as all the keys are known to the colluding users (thus $H[\mathcal{V}\backslash\{\left(v_k\right)_{k \in\mathcal{T}}\}]$ is trivially not connected). On the other hand, when the induced subgraph of $H$ is connected no matter which colluding user set is removed, we may fully use all existing groupwise keys to produce a secure summation scheme that is secure to all colluding user sets. The detailed proof of Theorem \ref{thm:feasible} is presented in Section \ref{sec:feasible}.

\section{Proof of Theorem \ref{thm:coded}}\label{pf:coded}
\subsection{Converse}
We start with a few useful lemmas. 
First, we show that each $X_k$ must contain at least $L$ symbols (the size of the input) even if all other inputs are known.
\begin{lemma}\label{lemma1}
For any $u \in [K]$, we have
\begin{eqnarray}
    H\left(X_u|(W_k,Z_k)_{k\in[K]\backslash \{u\}}\right) \geq L. \label{lemma1_eq}
\end{eqnarray}
\end{lemma}

\proof
\begin{eqnarray}
    && H\left(X_u|(W_k,Z_k)_{k\in[K]\backslash \{u\}}\right)\notag\\
    &\geq& I\left(X_u; \sum_{k\in[K]} W_k \Bigg|(W_k,Z_k)_{k\in[K]\backslash \{u\}}\right)\\
    &=& H\left(\sum_{k\in[K]} W_k \Bigg| (W_k,Z_k)_{k\in[K]\backslash \{u\}}\right) - H\left(\sum_{k\in[K]} W_k \Bigg| X_u, (W_k,Z_k)_{k\in[K]\backslash \{u\}}\right)\\
    &\overset{(\ref{ind})(\ref{message})}{\geq}& H\left(W_u\right) - 
    H\left(\sum_{k\in[K]} W_k \Bigg| (X_k)_{k\in[K]}\right)
    \label{pf_lemma1_1}\\
    &\overset{(\ref{h2})(\ref{corr})}{=}& L \label{eq:e1} 
\end{eqnarray}
where the first term of (\ref{pf_lemma1_1}) follows from the fact that input $W_u$ is independent of other inputs and keys $(W_k,Z_k)_{k \in [K]\backslash\{u\}}$ (see (\ref{ind})) and the second term of (\ref{pf_lemma1_1}) follows from the fact that $(X_k)_{k\in[K]\backslash \{u\}}$ is determined by $(W_k,Z_k)_{k\in[K]\backslash \{u\}}$ (see (\ref{message})). In (\ref{eq:e1}), we use the property that $W_u$ has $L$ uniform symbols (see (\ref{h2})) and the desired sum $\sum_{k\in[K]} W_k$ can be decoded with no error from all messages $(X_k)_{k\in[K]}$ (see (\ref{corr})).

\hfill\QED

Second, we show that the messages from any set of users must contain all information about their input sum (i.e., $L$ symbols).
\begin{lemma}\label{lemma3}
For any set of colluding users $\mathcal{T}$, 
denote its complement as $\mathcal{T}^c\triangleq[K]\backslash\mathcal{T}$ and we have
\begin{eqnarray}
    I\left((X_k)_{k \in \mathcal{T}^c};(W_k)_{k \in \mathcal{T}^c}|(W_k,Z_k)_{k \in \mathcal{T}}\right) =L.
    \label{lemma3_eq}
\end{eqnarray}
\end{lemma}

\proof
\begin{eqnarray}
    &&I\left((X_k)_{k \in \mathcal{T}^c};(W_k)_{k \in \mathcal{T}^c}|(W_k,Z_k)_{k \in \mathcal{T}}\right) \notag \\
    &=& I\left((X_k)_{k \in \mathcal{T}^c};(W_k)_{k \in \mathcal{T}^c}, \sum_{k \in \mathcal{T}^c} W_k \Bigg|(W_k,Z_k)_{k \in \mathcal{T}}\right)\\
    &=& I\left((X_k)_{k \in \mathcal{T}^c}; \sum_{k \in \mathcal{T}^c} W_k \Bigg|(W_k,Z_k)_{k \in \mathcal{T}}\right) + \underbrace{I\left((X_k)_{k \in \mathcal{T}^c};(W_k)_{k \in \mathcal{T}^c} \Bigg| \sum_{k \in \mathcal{T}^c} W_k, (W_k,Z_k)_{k \in \mathcal{T}}\right)}_{\overset{(\ref{security})}{=} 0} \notag\\
    &&\\
    &=& H\left( \sum_{k \in \mathcal{T}^c} W_k \Bigg|(W_k,Z_k)_{k \in \mathcal{T}}\right) - \underbrace{H\left(\sum_{k \in \mathcal{T}^c} W_k \Bigg|(X_k)_{k \in \mathcal{T}^c}, (W_k,Z_k)_{k \in \mathcal{T}}\right)}_{\overset{(\ref{message})(\ref{corr})}{=}0} \\
    &\overset{(\ref{ind})(\ref{h2})}{=}& L
\end{eqnarray}
where the last step follows from the independence and uniformity of the inputs.

\hfill\QED

Third, we show that the keys only known to non-colluding users 
must be sufficient large 
to protect their inputs (beyond the desired sum).
\begin{lemma}\label{lemma4}
For any set of colluding users $\mathcal{T}$ and its complement $\mathcal{T}^c = [K]\backslash\mathcal{T}$, we have
\begin{eqnarray}
    H((Z_k)_{k\in \mathcal{T}^c}|(Z_k)_{k\in \mathcal{T}})\geq (K-|\mathcal{T}|-1)L. \label{lemma4_eq}
\end{eqnarray}
\end{lemma}
\proof
First, 
we use Lemma {\ref{lemma1}} to obtain
\begin{eqnarray}
    &&H\left((X_k)_{k \in \mathcal{T}^c}|(W_k)_{k \in \mathcal{T}},(Z_k)_{k \in \mathcal{T}}\right)\notag\\
    &\geq& \sum_{u \in \mathcal{T}^c} H\left(X_u|(W_k,Z_k)_{k \in [K]\backslash\{u\}}\right)\label{pf_lemma4_1}\\
    &\overset{(\ref{lemma1_eq})}{\geq}& (K-|\mathcal{T}|)L \label{pf_lemma4_2}
\end{eqnarray}
where in (\ref{pf_lemma4_1}), we use the chain rule and the property that conditioning cannot increase entropy. 
Next,
\begin{eqnarray}
    && H((Z_k)_{k\in \mathcal{T}^c}|(Z_k)_{k\in \mathcal{T}})\notag\\
    &\geq& I\left((Z_k)_{k\in \mathcal{T}^c} ; (X_k)_{k \in \mathcal{T}^c} | (W_k)_{k \in \mathcal{T}^c}, (Z_k)_{k \in \mathcal{T}}\right)\\
    &\overset{(\ref{message})}{=}& H\left((X_k)_{k \in \mathcal{T}^c} | (W_k)_{k \in \mathcal{T}^c}, (Z_k)_{k \in \mathcal{T}}\right) \\ 
    &\geq& H\left((X_k)_{k \in \mathcal{T}^c} | (W_k)_{k \in [K]},(Z_k)_{k \in \mathcal{T}}\right)\\
    &=& H\left((X_k)_{k \in \mathcal{T}^c}|(W_k)_{k \in \mathcal{T}},(Z_k)_{k \in \mathcal{T}}\right) - I\left((X_k)_{k \in \mathcal{T}^c};(W_k)_{k \in \mathcal{T}^c}|(W_k)_{k \in \mathcal{T}},(Z_k)_{k \in \mathcal{T}}\right)\\
    &\overset{(\ref{pf_lemma4_2})(\ref{lemma3_eq})}{\geq}& (K-|\mathcal{T}|)L-L = (K-|\mathcal{T}|-1)L.
\end{eqnarray}

\hfill\QED

We are now ready to prove the converse of Theorem \ref{thm:coded}.

{\em Proof of $R \geq 1$:}
Consider any user $u \in [K]$.
\begin{eqnarray}
    L_X &\geq& H\left(X_u\right) \\
    &\geq& H\left(X_u|(W_k,Z_k)_{k\in[K]\backslash \{u\}}\right) \\
    &\overset{(\ref{lemma1_eq})}{\geq}& L \label{eq:e2} \\
    \Rightarrow ~~ R \overset{(\ref{rate:R})}{=} \frac{L_X}{L} &\geq& 1.
\end{eqnarray}

{\em Proof of $R_{Z} \geq 1$:}
First, we show that the message $X_u, u \in [K]$ is independent of the input $W_u$. 
\begin{eqnarray}
    &&I\left(X_u;W_u\right)\notag\\
    &\leq& I\left(X_u, \sum_{k\in [K]} W_k ;W_u\right)\\
    &=& I\left(\sum_{k\in [K]} W_k ;W_u\right) + I\left(X_u;W_u \Bigg| \sum_{k\in [K]} W_k \right)\label{pf_lemma2_1}\\
    &\overset{(\ref{ind})(\ref{h2})}{\leq}& I\left(\left(W_k\right)_{k\in[K]}; \left(X_k\right)_{k\in[K]} \Bigg| \sum_{k\in [K]} W_k \right)\label{pf_lemma2_2}\\
    &\overset{(\ref{security})}{=}& 0\label{pf_thm1_3}
\end{eqnarray}
where (\ref{pf_lemma2_2}) follows from the observation that the first term of (\ref{pf_lemma2_1}) is zero because $K \geq 2$ and the inputs are independent and uniform. To obtain (\ref{pf_thm1_3}), we set $\mathcal{T} = \emptyset$ in the security constraint (\ref{security}).
Next, consider any $u \in [K]$.
\begin{eqnarray}
    L_Z &\geq& H\left(Z_u\right)\\
    &\geq& I\left(Z_u;X_u|W_u\right) \\
    &\overset{(\ref{message})}{=}& H\left(X_u|W_u\right) \\ 
    &\overset{(\ref{pf_thm1_3})}{=}& H\left(X_u\right)\\ 
    &\overset{(\ref{eq:e2})}{\geq}& L \\
    \Rightarrow ~~ R_{Z} \overset{(\ref{rate:R_Z})}{=} \frac{L_Z}{L} &\geq& 1.
\end{eqnarray}

{\em Proof of $R_{Z_{\Sigma}} \geq K-1$:}
Set $\mathcal{T} = \emptyset$ so that its complement $\mathcal{T}^c = [K]$.
\begin{eqnarray}
    L_{Z_{\Sigma}} &\geq& H(Z_{\Sigma})\\
    &\overset{(\ref{total rand})}{\geq}& H\left((Z_k)_{k \in [K]}\right)\label{pf_thm1_4}\\
    &\geq& H((Z_k)_{k\in \mathcal{T}^c}|(Z_k)_{k\in \mathcal{T}})\\
    &\overset{(\ref{lemma4_eq})}{\geq}& (K-|\mathcal{T}|-1)L = (K-1)L\label{pf_thm1_5}\\
    \Rightarrow ~~ R_{Z_{\Sigma}} \overset{(\ref{rate:R_Z})}{=} \frac{L_{Z_{\Sigma}}}{L} &\geq& K-1
\end{eqnarray}
where in (\ref{pf_thm1_4}), we use the fact that all key variables $(Z_k)_{k \in [K]}$ are generated from the source key variable $Z_{\Sigma}$ (see (\ref{total rand})). In (\ref{pf_thm1_5}), we use (\ref{lemma4_eq}) in Lemma {\ref{lemma4}}. 

\subsection{Achievability}\label{sec:ach1}
The achievable scheme is straightforward and is plotted in Fig.~\ref{fig:model}. The only non-trivial aspect is the proof of security, presented below.

We first describe the scheme.
Consider $K-1$ i.i.d. uniform variables over $\mathbb{F}_q$, $N_1,\cdots,N_{K-1}$ and set the key variables as 
 \begin{eqnarray}
     Z_k &=& N_k, \forall k \in [K-1] \notag\\
     Z_K &=& -N_1 - \cdots - N_{K-1} \triangleq N_K. \label{thm1_ind}
 \end{eqnarray}
 Set $L=1$ and set the messages as
 \begin{eqnarray}
     X_k = W_k + N_k, \forall k \in [K].\label{thm1_message}
 \end{eqnarray}
 Note that $L_X = 1, L_Z = 1, L_{Z_\Sigma} = K-1$, so the rate achieved is $R = L_X/L = 1, R_Z = L_Z/L = 1, R_{Z_\Sigma} = L_{Z_\Sigma}/L = K-1$, as desired. Correctness is proved by noting that $\sum_{k\in[K]} {X_k} = \sum_{k\in [K]} W_k$. We are left to prove the security. For any colluding user set $\mathcal{T}$, we verify that the security constraint (\ref{security}) is satisfied.
 \begin{eqnarray}
     && I\left(\left(W_k\right)_{k \in [K]}; \left(X_k\right)_{k \in [K]} \Bigg| \sum_{k \in [K]} W_k, \left(W_k, Z_k\right)_{k \in \mathcal{T}} \right)\notag\\
     &\overset{(\ref{thm1_ind}) (\ref{thm1_message})}{=}& I\left(\left(W_k\right)_{k \in \mathcal{T}^c}; \left(W_k + N_k\right)_{k \in \mathcal{T}^c} \Bigg| \sum_{k \in \mathcal{T}^c} W_k, \left(W_k, N_k\right)_{k \in \mathcal{T}} \right)\label{eq_thm1_a2}\\
     &=& H\left(\left(W_k+N_k\right)_{k \in \mathcal{T}^c}\Bigg| \sum_{k \in \mathcal{T}^c} W_k, \left(W_k, N_k\right)_{k \in \mathcal{T}} \right)\notag\\
     &&-~ H\left(\left(W_k+N_k\right)_{k \in \mathcal{T}^c} \Bigg| \sum_{k \in \mathcal{T}^c} W_k, \left(W_k, N_k\right)_{k \in \mathcal{T}}, \left(W_k\right)_{k \in \mathcal{T}^c} \right)\\
     &\leq& H\left(\left(W_k+N_k\right)_{k \in \mathcal{T}^c}\Bigg| \sum_{k \in \mathcal{T}^c} (W_k+N_k) \right) - H\left(\left(N_k\right)_{k \in \mathcal{T}^c} | \left(N_k\right)_{k \in \mathcal{T}}\right)\label{eq_thm1_a3}\\
     &\leq& (K-|\mathcal{T}|-1) - (K-|\mathcal{T}|-1) = 0
\end{eqnarray}
 where in (\ref{eq_thm1_a2}), we plug in the message and key variable assignment (refer to (\ref{thm1_ind}), (\ref{thm1_message})). In (\ref{eq_thm1_a3}), the first term follows from fact that $0 = \sum_{k\in[K]} N_k = \sum_{k\in \mathcal{T}} N_k + \sum_{k\in \mathcal{T}^c} N_k$ (refer to (\ref{thm1_ind})), i.e., $\sum_{k\in \mathcal{T}^c} N_k$ can be obtained from $(N_k)_{k\in\mathcal{T}}$ and the property that dropping conditioning cannot reduce entropy; the second term is obtained by applying the independence of the input and key variables (refer to (\ref{ind})). In the last step, the first term follows from the fact that the first term contains at most $K-|\mathcal{T}|-1$ terms after conditioning and uniform distribution maximizes entropy; the second term follows from the independence and uniformity of the $N_1, \cdots, N_{K-1}$ variables and the fact that $\sum_{k\in \mathcal{T}^c} N_k$ can be obtained from $\left(N_k\right)_{k \in \mathcal{T}}$.

\section{Proof of Theorem \ref{thm:groupwise}}\label{sec:groupwise}
\subsection{Converse}
When $G > K-T$, consider any set of colluding users $\mathcal{T}$, where $|\mathcal{T}| = T \leq K-2$. Each groupwise key $S_{\mathcal{G}}, \forall \mathcal{G} \subset [K], |\mathcal{G}| = G$ is known to the colluding users so that the key hypergraph has no edges and has at least two nodes after removing the colluding users and we may invoke Theorem \ref{thm:feasible} to establish that $\mathcal{R}_g^* = \emptyset$. Henceforth, it suffices to consider the case when $G \leq K-T$.

{\em Proof of $R\geq 1$}: It follows from the proof of Theorem \ref{thm:coded} as groupwise keys are a special case of arbitrarily coded keys.

{\em Proof of $R_S \geq \frac{K-T-1}{\binom{K-T}{V}}$:} Note that Lemma \ref{lemma4} holds for groupwise keys. 
Applying (\ref{lemma4_eq}) to a colluding user set $\mathcal{T}$ such that $|\mathcal{T}| = T$, we have
\begin{eqnarray}
(K-T-1)L &\overset{(\ref{lemma4_eq})}{\leq}& H\left((Z_k)_{k\in \mathcal{T}^c}|(Z_k)_{k\in \mathcal{T}}\right)\\
    &\overset{(\ref{Z_individual})}{=}& H\left((S_\mathcal{G})_{\mathcal{G} \in \binom{[K]}{G},\mathcal{G} \cap \mathcal{T}^c \neq \emptyset} \Big| (S_\mathcal{G})_{\mathcal{G} \in \binom{[K]}{G},\mathcal{G} \cap \mathcal{T}\neq \emptyset}\right) \label{eq:e3}\\
    &\overset{(\ref{ind_S})}{=}& H\left((S_\mathcal{G})_{\mathcal{G} \in \binom{[K]}{G},\mathcal{G} \cap \mathcal{T}^c \neq \emptyset,\mathcal{G} \cap \mathcal{T} = \emptyset}\right) \label{eq:e4}\\
    &=& H\left((S_\mathcal{G})_{\mathcal{G} \in \binom{\mathcal{T}^c}{G}}\right)\\
    &\overset{(\ref{ind_S})}{=}& \binom{K-T}{G} \times L_S\\
    \Rightarrow ~~ R_S\overset{(\ref{rate:R_S})}{=} \frac{L_S}{L} &\geq& \cfrac{K-T-1}{\binom{K-T}{G}} \label{eq:e5}
\end{eqnarray}
where in (\ref{eq:e3}), we replace each key variable $Z_k$ by the groupwise keys (refer to (\ref{Z_individual})) and in (\ref{eq:e4}) and (\ref{eq:e5}), we apply the independence of the groupwise keys (refer to (\ref{ind_S})).

\subsection{Achievability}\label{section:thm2_achi}
We first present the achievable scheme for two examples, to illustrate the idea in a simpler setting.

\subsubsection{Example: $K=3,T=0,G=2$}\label{Example1}
Consider $K=3$ users, where no user may collude with the server $(T=0)$. As the group size $G=2$, we have 3 groupwise keys $S_{\{1,2\}},S_{\{1,3\}},S_{\{2,3\}}$. 

The achievable scheme is based on interpreting the scheme from Section \ref{sec:ach1} of Theorem \ref{thm:coded} in the groupwise key setting and permuting it for symmetrization. Specifically, consider the following basic component of the secure summation scheme from Theorem \ref{thm:coded}.
\begin{eqnarray}
X_1 &=& W_1 + A \\
X_2 &=& W_2 + B \\
X_3 &=& W_3 - A - B
\end{eqnarray}
where each $X_k, W_k$ contains $1$ symbol from $\mathbb{F}_q$ and $A, B$ are two i.i.d. uniform key symbols from $\mathbb{F}_q$. Note that the above keys can be assigned through setting groupwise keys $S_{\{1,3\}} = A, S_{\{2,3\}} = B$. Correctness and security follow immediately from Theorem \ref{thm:coded}. To produce a symmetric scheme where each groupwise key has the same size (note that $S_{\{1,2\}}$ has not been used), we apply the above scheme to all permutation of the users $\{1,2,3\}$. In particular, we set $L = 3! = 6$ and each input has $6$ symbols, i.e., $W_k = [W_k(1), \cdots, W_k(6)]^T$. Correspondingly, uniform key variables $A, B$ have length $6$ each. The message and key assignment is as follows.
\begin{align}
\begin{array}{ccc}
    \begin{array}{l}
        X_1(1) = W_1(1) + A(1)\\
        X_2(1) = W_2(1) + B(1)\\
        X_3(1) = W_3(1) - A(1)-B(1)
    \end{array}
     &
     \begin{array}{l}
        X_1(2) = W_1(2) + B(2)\\
        X_2(2) = W_2(2) - A(2)-B(2)\\
        X_3(2) = W_3(2) + A(2)
    \end{array}
     &
     \begin{array}{l}
        X_1(3) = W_1(3) - A(3)-B(3)\\
        X_2(3) = W_2(3) + A(3)\\
        X_3(3) = W_3(3) + B(3)
    \end{array}
\end{array}\notag\\
\begin{array}{ccc}
    \begin{array}{l}
        X_1(4) = W_1(4) + A(4)\\
        X_2(4) = W_2(4) - A(4)-B(4)\\
        X_3(4) = W_3(4) + B(4)
    \end{array}
     &
     \begin{array}{l}
        X_1(5) = W_1(5) + B(5)\\
        X_2(5) = W_2(5) + A(5)\\
        X_3(5) = W_3(5) - A(5)-B(5)
    \end{array}
     &
     \begin{array}{l}
        X_1(6) = W_1(6) - A(6)-B(6)\\
        X_2(6) = W_2(6) + B(6)\\
        X_3(6) = W_3(6) + A(6)
    \end{array}
\end{array}\notag
\end{align}
and
\begin{eqnarray}
    S_{\{1,2\}} = \left[
    \begin{array}{c}
        B(2)\\
        A(3)\\
        A(4)\\
        B(6)
    \end{array}\right]
    &
    S_{\{1,3\}} = \left[
    \begin{array}{c}
        A(1)\\
        B(3)\\
        B(5)\\
        A(6)
    \end{array}\right]
    &
    S_{\{2,3\}} = \left[
    \begin{array}{c}
        B(1)\\
        A(2)\\
        B(4)\\
        A(5)
    \end{array}\right]
\end{eqnarray}
where $L_X = 6, L_S = 4$ so that the rate achieved is $R = L_X/L = 1, R_S = L_S/L = 2/3 = (3-1)/\binom{3}{2}$, as desired. 

Permutation retains the correctness and security of the scheme (note that the input and key variables for each permutation are independent so that the mutual information terms (\ref{eq_thm1_a2}) in the security proof tensorize).

To facilitate the presentation of the achievable scheme when $T > 0$, it is convenient to describe the scheme in matrix form. In particular, the above scheme can be equivalently written as 
\begin{eqnarray}\label{example1_message}
\left[
    \begin{array}{c}
         X_1\\
         X_2\\
         X_3
    \end{array}\right]=\left[
    \begin{array}{c}
         W_1\\
         W_2\\
         W_3
    \end{array}\right]+
    \underbrace{\left[
    \begin{array}{ccc}
        {\bf H}_{\{1,2\}} & {\bf H}_{\{1,3\}} & {\bf 0}_{6 \times 4}\\
        -{\bf H}_{\{1,2\}} & {\bf 0}_{6 \times 4} & {\bf H}_{\{2,3\}}\\
        {\bf 0}_{6 \times 4} & -{\bf H}_{\{1,3\}} & -{\bf H}_{\{2,3\}}\\
    \end{array}\right]}_{\triangleq{\bf H}_{18 \times 12}}
    \left[\begin{array}{c}
         S_{\{1,2\}}\\
         S_{\{1,3\}}\\
         S_{\{2,3\}}
    \end{array}
    \right]
\end{eqnarray}
where
\begin{eqnarray}\label{example1_H}
    {\bf H}_{\{1,2\}}=\left[
    \begin{array}{cccc}
         0 & 0 & 0 & 0\\
         1 & 0 & 0 & 0\\
         0 & -1 & 0 & 0\\
         0 & 0 & 1 & 0\\
         0 & 0 & 0 & 0\\
         0 & 0 & 0 & -1
    \end{array}\right] ~
    {\bf H}_{\{1,3\}}=\left[
    \begin{array}{cccc}
         1 & 0 & 0 & 0\\
         0 & 0 & 0 & 0\\
         0 & -1 & 0 & 0\\
         0 & 0 & 0 & 0\\
         0 & 0 & 1 & 0\\
         0 & 0 & 0 & -1
    \end{array}\right] ~
    {\bf H}_{\{2,3\}}=\left[
    \begin{array}{cccc}
         1 & 0 & 0 & 0\\
         0 & -1 & 0 & 0\\
         0 & 0 & 0 & 0\\
         0 & 0 & -1 & 0\\
         0 & 0 & 0 & 1\\
         0 & 0 & 0 & 0
    \end{array}\right].
\end{eqnarray}
Now the scheme design reduces to the assignment of the key precoding matrix ${\bf H}$ and it turns out that security can be guaranteed by certain rank property of ${\bf H}$ (refer to Lemma \ref{lemma:rank}).

\subsubsection{Example: $K=5,T=2,G=2$}
Consider $K=5$ users, where at most $T=2$ users may collude with the server, and every $G=2$ users share a groupwise key. The rate to be achieved is $R = 1, R_S=(K-T-1)/\binom{K-T}{G}=2/3$. To this end, set $L=L_X = 3$, i.e., $W_k=[W_k(1),W_k(2),W_k(3))]^T \in \mathbb{F}_q^{3\times 1}, k \in [3]$; set $L_S=2$, i.e., $S_\mathcal{G}=[S_\mathcal{G}(1),S_\mathcal{G}(2)]^T \in \mathbb{F}_q^{2\times 1}, \mathcal{G} \subset [3],|\mathcal{G}|=2$. 
The messages are set as
\begin{align}
     X_1 &= W_1 +{\bf H}_{\{1,2\}}^{3\times 2}S_{\{1,2\}} +{\bf H}_{\{1,3\}}S_{\{1,3\}}+{\bf H}_{\{1,4\}}S_{\{1,4\}}+{\bf H}_{\{1,5\}}S_{\{1,5\}} \notag\\
     X_2 &= W_2-{\bf H}_{\{1,2\}}S_{\{1,2\}}+{\bf H}_{\{2,3\}}S_{\{2,3\}}+{\bf H}_{\{2,4\}}S_{\{2,4\}}+{\bf H}_{\{2,5\}}S_{\{2,5\}} \notag\\
     X_3 &= W_3-{\bf H}_{\{1,3\}}S_{\{1,3\}}-{\bf H}_{\{2,3\}}S_{\{2,3\}}+{\bf H}_{\{3,4\}}S_{\{3,4\}}+{\bf H}_{\{3,5\}}S_{\{3,5\}} \notag\\
     X_4 &= W_4-{\bf H}_{\{1,4\}}S_{\{1,4\}}-{\bf H}_{\{2,4\}}S_{\{2,4\}}-{\bf H}_{\{3,4\}}S_{\{3,4\}}+{\bf H}_{\{4,5\}}S_{\{4,5\}} \notag\\
     X_5 &= W_5-{\bf H}_{\{1,5\}}S_{\{1,5\}}-{\bf H}_{\{2,5\}}S_{\{2,5\}}-{\bf H}_{\{3,5\}}S_{\{3,5\}}-{\bf H}_{\{4,5\}}S_{\{4,5\}} \label{eq:hd}
\end{align}
where each ${\bf H}_{\mathcal{G}}\in \mathbb{F}_q^{3 \times 2}$ is a $3 \times 2$ key precoding matrix. Note that ${\bf H}_{\mathcal{G}}$ and $-{\bf H}_{\mathcal{G}}$ are used by every $|\mathcal{G}|=2$ users (i.e., zero-sum-randomness) so that correctness is guaranteed, i.e., $\sum_{k\in[5]} X_k = \sum_{k\in[5]} W_k$. Security is ensured if the matrices ${\bf H}_{\mathcal{G}}$ are sufficiently generic. In particular, we will show in the general proof that if ${\bf H}_{\mathcal{G}}$ are randomly drawn from a sufficiently large field (note that the size $q$ of the field $\mathbb{F}_q$ remains the same, but we can code over longer blocks by enlarging input size $L$ so that we are operating over a larger extension field), then there must exist a matrix construction such that security holds. For this setting, suppose $q = 5$ and we may set
\begin{align}
     {\bf H}_{\{1,2\}}= \left[
     \begin{array}{cc}
          3&3  \\
          1&4\\
          2&4
     \end{array}\right]
     {\bf H}_{\{1,3\}}= \left[
     \begin{array}{cc}
          2&1  \\
          0&4\\
          0&1
     \end{array}\right]~
     {\bf H}_{\{1,4\}}= \left[
     \begin{array}{cc}
          4&1  \\
          1&0\\
          4&1
     \end{array}\right]~
     {\bf H}_{\{1,5\}}= \left[
     \begin{array}{cc}
          3&4  \\
          2&2\\
          1&2
     \end{array}\right]~
     {\bf H}_{\{2,3\}}= \left[
     \begin{array}{cc}
          4&3  \\
          1&1\\
          3&2
     \end{array}\right]~\notag\\
     {\bf H}_{\{2,4\}}= \left[
     \begin{array}{cc}
          0&3  \\
          0&4\\
          2&0
     \end{array}\right]~
     {\bf H}_{\{2,5\}}= \left[
     \begin{array}{cc}
          2&1  \\
          2&0\\
          0&3
    \end{array}\right]~
     {\bf H}_{\{3,4\}}= \left[
     \begin{array}{cc}
          1&3  \\
          2&1\\
          0&3
     \end{array}\right]~
     {\bf H}_{\{3,5\}}= \left[
     \begin{array}{cc}
          3&0  \\
          3&1\\
          2&4
     \end{array}\right]~
     {\bf H}_{\{4,5\}}= \left[
     \begin{array}{cc}
          0&4  \\
          4&0\\
          2&2
     \end{array}\right].\label{V_example2}
 \end{align}
Let us see now why security is guaranteed. Intuitively, we require the key variables to fully cover the messages, which will translate to the requirement that certain matrices have full rank.
Suppose we have $|\mathcal{T}| = 2$ colluding users, say $\mathcal{T} = \{4,5\}$. Consider the security constraint (\ref{security}).
\begin{eqnarray}
     &&I\left(W_1,W_2,W_3;X_1,X_2,X_3|W_1+W_2+W_3,W_4,W_5,Z_4,Z_5\right) \notag\\
     &=& H\left(X_1,X_2,X_3|W_1+W_2+W_3,W_4,W_5,Z_4,Z_5\right)\notag\\
     && -~H\left(X_1,X_2,X_3|W_1,W_2,W_3,W_4,W_5,Z_4,Z_5\right) \\
     &\overset{(\ref{eq:hd})}{\leq}& 2L - H\left( \underbrace{\left[
    \begin{array}{cccccccccc}
        {\bf H}_{\{1,2\}}& {\bf H}_{\{1,3\}} & {\bf 0} \\
        -{\bf H}_{\{1,2\}}& {\bf 0}& {\bf H}_{\{2,3\}} \\
        {\bf 0}& -{\bf H}_{\{1,3\}} &- {\bf H}_{\{2,3\}} 
    \end{array}\right]}_{= \hat{\bf H}}
    \left[ \begin{array}{c}
         S_{\{1,2\}}  \\
         S_{\{1,3\}} \\
         S_{\{2,3\}}
    \end{array}\right]
     \right) \label{eq:s2} \\
     &=& 2L ~-H\left(S_{\{1,2\}},S_{\{1,3\}},S_{\{2,3\}}\right)\label{eq_example2_2}\\
     &=& 2 L -3 L_S = 2\times3 -3\times2=0
\end{eqnarray}
where in (\ref{eq:s2}), the first term follows from the maximum number of symbols contained in $X_1, X_2, X_3$ after conditioning on $X_1+X_2+X_3 = W_1+W_2+W_3$; the second term follows from the design and the independence of the keys and messages (\ref{eq:hd}). To obtain (\ref{eq_example2_2}), we require $\hat{\bf H}$ to have full rank so that the precoded keys are invertible to the original groupwise keys (i.e., the key precoding matrices are sufficiently generic).
We may readily verify that $\hat{\bf H}$ has full rank for the assignment (\ref{V_example2}). Note that $\hat{\bf H}$ has the same form as (\ref{example1_message}) in the $T=0$ case so that the assignment in (\ref{example1_message}) can also be used here (such a reduction will be used in the general proof). Therefore, we have seen how the security proof can be translated to the full rank property of the precoding matrices (see Lemma \ref{lemma:rank} for the general result). The security for other cases of colluding users can be similarly verified.

We are now ready to proceed to the general proof. As $\mathcal{R}_g^* = \emptyset$ when $G > K-T$, we only need to consider the case where $G \leq K-T$.

\subsubsection{General Proof for Arbitrary $K, T, G \leq K-T$}
Suppose\footnote{The input and key lengths are set to be larger than those in the examples to simplify the proof. In particular, $K!$ is introduced to allow permutation as in Section \ref{Example1} to produce matrices with desired ranks. $M$ is an integer scaling factor to work over the (larger) extension field.} $L=L_X=K!\binom{K-T}{G}M$ and $L_S = K!(K-T-1)M$ so that the desired rate is achieved. 
Group every $M$ symbols from $W_k$ together and view them as a single symbol from the extension field $\mathbb{F}_{q^M}$, i.e., $W_k \in \mathbb{F}_{q^M}^{L/M \times 1}$. Similarly, suppose $S_{\mathcal{G}} \in \mathbb{F}_{q^M}^{L_S/M \times 1}$.
The messages are set as 
\begin{eqnarray}
    X_k &=& W_k + \sum_{{\mathcal{G}:k \in \mathcal{G},\mathcal{G}\in \binom{[K]}{G}}} {\bf H}_{\mathcal{G}}^{k}S_{\mathcal{G}},~\forall k\in[K] \label{eq:sc}
\end{eqnarray}
where ${\bf H}_{\mathcal{G}}^{k}$ is an {\color{black} $L/M\times L_S/M$}
matrix over $\mathbb{F}_{q^M}$ such that for any group, the sum of the precoded keys is zero, i.e.,
\begin{eqnarray}\label{Vsum}
    \sum_{k \in \mathcal{G}}{\bf H}_{\mathcal{G}}^k={\bf 0}_{L/M\times L_S/M},~\forall \mathcal{G}\in \binom{[K]}{G}. \label{eq:zerosum}
\end{eqnarray}
Collecting all messages and writing them in a matrix form, we have
\begin{eqnarray}
\left[
    \begin{array}{c}
         X_1\\
         \vdots\\
         X_K
    \end{array}\right]=\left[
    \begin{array}{c}
         W_1\\
         \vdots\\
         W_K
    \end{array}\right]+ {\bf H}
    \left[\begin{array}{c}
         S_{\{1,2,\cdots,G\}}\\
         \vdots\\
         S_{\{K-G+1,\cdots,K\}}
    \end{array}
    \right]
\end{eqnarray}
where
\begin{eqnarray}
    {\bf H} = \left[ \left({\bf H}_{\mathcal{G}}^k\right)_{k\in[K], \mathcal{G} \in \binom{[K]}{G}} \right]
    \triangleq \left[\begin{array}{cccc}
        {\bf H}_{\{1,2,\cdots,G\}}^1  &\cdots& {\bf H}_{\{K-G+1,\cdots,K\}}^1\\
        \vdots  & \ddots &\vdots\\
        {\bf H}_{\{1,2,\cdots,G\}}^K &\cdots& {\bf H}_{\{K-G+1,\cdots,K\}}^K\\
    \end{array}\right] \label{eq:h}
\end{eqnarray}
and ${\bf H}_{\mathcal{G}}^k={\bf 0}_{L/M\times L_S/M}$,
$\forall k \in [K], \mathcal{G} \in \binom{[K]}{G}, k \notin \mathcal{G}$.

Correctness is straightforward, as
\begin{eqnarray}
    \sum_{k \in [K]} X_k &=& \sum_{k \in [K]} W_k + \sum_{k\in[K]} \sum_{{\mathcal{G}:k \in \mathcal{G},\mathcal{G}\in \binom{[K]}{G}}} {\bf H}_{\mathcal{G}}^{k}S_{\mathcal{G}}\\ 
    &=& \sum_{k \in [K]}W_k +\sum_{\mathcal{G}:k\in\mathcal{G},\mathcal{G}\in\binom{[K]}{G}} \left[\left(\sum_{k \in \mathcal{G}}{\bf H}_{\mathcal{G}}^k\right)S_{\mathcal{G}}\right]\\
    &\overset{(\ref{Vsum})}{=}& \sum_{k \in [K]} W_k.
\end{eqnarray}

We show that security is guaranteed if some precoding matrix has certain rank, presented in the following lemma. The proof of the existence of such matrices is deferred to the next section.

Consider any colluding user set $\mathcal{T}$. Define the following submatrix of ${\bf H}$ (which is obtained by considering the keys only known to non-colluding users).
\begin{eqnarray}
\hat{\bf H}[\mathcal{T}] \triangleq \left[ \left({\bf H}_{\mathcal{G}}^k\right)_{k\in[K]\backslash\mathcal{T}, \mathcal{G} \in \binom{[K]\backslash\mathcal{T}}{G}} \right] \label{eq:hhat}
\end{eqnarray}
which contains $K-|\mathcal{T}|$ row blocks and $\binom{K-|\mathcal{T}|}{G}$ column blocks of ${\bf H}_{\mathcal{G}}^k$ terms.

\begin{lemma}\label{lemma:rank}
For any colluding user set $\mathcal{T} \subset [K], |\mathcal{T}|\leq T$, the scheme (\ref{eq:sc}) satisfies the security constraint (\ref{security}) if and only if
$\mbox{rank}(\hat{\bf H}[\mathcal{T}]) = (K-|\mathcal{T}|-1)L/M$ over $\mathbb{F}_{q^M}$.
\end{lemma}
\proof
Consider the `if' direction. Consider the security constraint (\ref{security}).
\begin{eqnarray}
    &&I\left(\left(W_k\right)_{k\in[K]}; \left(X_k\right)_{k\in[K]} \Bigg| \sum_{k\in [K]} W_k, \left( W_k, Z_k \right)_{k\in\mathcal{T}} \right)\notag\\
    &=&I\left(\left(W_k\right)_{k\in[K]\backslash\mathcal{T}}; \left(X_k\right)_{k\in[K]\backslash\mathcal{T}}\Bigg|\sum_{k\in [K]\backslash\mathcal{T}} W_k, \left( W_k, Z_k \right)_{k\in\mathcal{T}}\right) \label{eq:s5}\\
    &=&H\left(\left(X_k\right)_{k\in[K]\backslash\mathcal{T}}\Bigg|\sum_{k\in [K]\backslash\mathcal{T}} W_k, \left( W_k, Z_k \right)_{k\in\mathcal{T}}\right)
    - H\left(\left(X_k\right)_{k\in[K]\backslash\mathcal{T}}\Bigg|(W_k)_{k\in[K]},\left( Z_k \right)_{k\in\mathcal{T}}\right) \notag\\
    &&\\
    &=& H\left( \left(X_k\right)_{k\in[K]\backslash\mathcal{T}} \Bigg|\sum_{k\in [K]\backslash\mathcal{T}} W_k, \left( W_k, Z_k \right)_{k\in\mathcal{T}}\right)\notag\\
    && ~- H\left(\left(
    \sum_{{\mathcal{G}:k \in \mathcal{G},\mathcal{G}\in \binom{[K]\backslash \mathcal{T}}{G}}} {\bf H}_{\mathcal{G}}^{k}S_{\mathcal{G}}
    \right)_{k\in[K]\backslash\mathcal{T}}\Bigg|(W_k)_{k\in[K]},\left( Z_k \right)_{k\in\mathcal{T}}\right)\label{lemma:rank_eq2}\\
    &\leq& (K-|\mathcal{T}|-1)L -H\left( \hat{\bf H}[\mathcal{T}] \left(S_{\mathcal{G}}\right)_{\mathcal{G}\in \binom{[K]\backslash \mathcal{T}}{G}} \right) \label{eq:ss}\\
    &=& (K-|\mathcal{T}|-1)L - (K-|\mathcal{T}|-1)L = 0
\end{eqnarray}
where in (\ref{eq:ss}), the first term follows from the fact that $\sum_{k\in[K]\backslash\mathcal{T}} X_k = \sum_{k\in[K]\backslash\mathcal{T}} W_k$ and uniform distribution maximizes entropy (note that the entropy is measured in $q$-ary units); the second term is obtained by using the definition of $\hat{\bf H}[\mathcal{T}]$ (\ref{eq:hhat}), the independence of the input and key variables (\ref{ind}), and the independence of the groupwise key variables (\ref{ind_S}).
In the last step, we use the assumption that $\hat{\bf H}[\mathcal{T}]$ has rank $(K-|\mathcal{T}|-1)L/M$ over $\mathbb{F}_{q^M}$ and $S_\mathcal{G}$ symbols are i.i.d. and uniform.

The `only if' direction is obvious. For any scheme of form (\ref{eq:sc}), if (\ref{eq:s5}) is zero, then all inequalities above must be strictly equality, i.e., $\hat{\bf H}[\mathcal{T}]$ must have rank $(K-|\mathcal{T}|-1)L/M$.

\hfill\QED

\subsubsection{Existence of ${\bf H}$: Reduction to $T=0$}\label{schemeT=0}
In this section, we show that there exists a matrix ${\bf H}$ of form (\ref{eq:h}) such that its submatrix $\hat{\bf H}[\mathcal{T}]$, specified in (\ref{eq:hhat}), has rank $(K-|\mathcal{T}|-1)L/M$ over $\mathbb{F}_{q^M}$ for all possible colluding user sets $\mathcal{T}, \forall \mathcal{T} \subset [K], |\mathcal{T}| \leq T$.

We show that if for each $\mathcal{G} \in \binom{[K]}{G}$, we generate each element of any $G-1$ matrices ${\bf H}_\mathcal{G}^k, k \in \mathcal{G}$ in (\ref{eq:h}) uniformly and i.i.d. over the extension field $\mathbb{F}_{q^M}$ (and the last matrix is set as the negative of the sum of the remaining $G-1$ matrices, to satisfy (\ref{eq:zerosum})), then as\footnote{Note that here we present an existence proof over a exceedingly large block lengths based on probabilistic arguments, similar to the Shannon's original random coding proof to channel capacity. We leave the problem of finding a short capacity achieving code as interesting future work.} $M \rightarrow \infty$, the probability that $\hat{\bf H}[\mathcal{T}]$ has rank $(K-|\mathcal{T}|-1)L/M$ approaches $1$ such that the existence of ${\bf H}$ is guaranteed. To apply the Schwartz–Zippel lemma \cite{Demillo_Lipton, Schwartz, Zippel}, we need to guarantee that for each $\mathcal{T}$, there exists a realization of $\hat{\bf H}[\mathcal{T}]$ such that $\mbox{rank}(\hat{\bf H}[\mathcal{T}]) = (K-|\mathcal{T}|-1)L/M$.

We are left to show that for any fixed $\mathcal{T}$, we have an $\hat{\bf H}[\mathcal{T}]$ of rank $(K-|\mathcal{T}|-1)L/M$ over $\mathbb{F}_{q^M}$.
Note that $\hat{\bf H}[\mathcal{T}]$ is the same as the precoding matrix ${\bf H}$ in (\ref{eq:h}) when we have $K - |\mathcal{T}|$ users in $[K]\backslash\mathcal{T}$, $0$ colluding users, and groupwise keys of group size $G$. Here from the proof of Theorem \ref{thm:coded}, we have a scalar ($L=1$) linear basic scheme of form (\ref{eq:sc}) and then following the example in Section \ref{Example1}, we can permute the basic scheme to produce a length $L=K!$ scheme with symmetric groupwise keys. We can further extend the field to $\mathbb{F}_{q^{M}}$ by considering a block of $M$ symbols together. Then by repeating the scheme $\binom{K-T}{G}$ times gives us the desired input length $L = K!\binom{K-T}{G} M$. Repeating and permuting the basic scheme preserves correctness and security. By the `only if' direction of Lemma \ref{lemma:rank}, such a secure scheme will produce the desired $\hat{\bf H}[\mathcal{T}]$.

\section{Proof of Theorem \ref{thm:feasible}}\label{sec:feasible}
The proof of Theorem \ref{thm:feasible} is comprised of two directions. On the one hand, we show that $\mathcal{R}_{gc}^* \neq \emptyset \Rightarrow H[\mathcal{V}\backslash\{\left(v_k\right)_{k \in\mathcal{T}}\}]~\mbox{is connected for any}~\mathcal{T} \in \overline{\mathcal{T}}$, i.e., $H[\mathcal{V}\backslash\{\left(v_k\right)_{k \in\mathcal{T}}\}]$ is not connected for some $\mathcal{T} \in \overline{\mathcal{T}} \Rightarrow \mathcal{R}_{gc}^* = \emptyset$, which requires a converse proof. On the other hand, we show that $\mathcal{R}_{gc}^* \neq \emptyset \Leftarrow H[\mathcal{V}\backslash\{\left(v_k\right)_{k \in\mathcal{T}}\}]~\mbox{is connected for any}~\mathcal{T} \in \overline{\mathcal{T}}$, which requires an achievability proof. These two proofs are provided next.

\subsection{Converse: $\exists \mathcal{T}, H[\mathcal{V}\backslash\{\left(v_k\right)_{k \in\mathcal{T}}\}]$ is not connected $\Rightarrow \mathcal{R}_{gc}^* = \emptyset$}

As $H[\mathcal{V}\backslash\{\left(v_k\right)_{k \in\mathcal{T}}\}]$ is not connected, by definition, there exists a two-part partition of the nodes, $\mathcal{V}_1, \mathcal{V}_2$, where $\mathcal{V}_1 \cup \mathcal{V}_2 = \mathcal{V}\backslash \{\left(v_k\right)_{k \in\mathcal{T}}\}, |\mathcal{V}_1|\geq 1, |\mathcal{V}_2|\geq 1$ such that any edge $e$ of $H[\mathcal{V}\backslash\{\left(v_k\right)_{k \in\mathcal{T}}\}]$ must fully lie in $\mathcal{V}_1$ or $\mathcal{V}_2$, i.e., $e \subset \mathcal{V}_1$ or $e \subset \mathcal{V}_2$. Denote the set of the indices of the nodes in $\mathcal{V}_1$ and $\mathcal{V}_2$ (i.e., the set of users) by $\mathcal{U}_1$ and $\mathcal{U}_2$, respectively. Then $\mathcal{U}_1 \cup \mathcal{U}_2 = [K]\backslash\mathcal{T}, |\mathcal{U}_1| \geq 1, |\mathcal{U}_2| \geq 1$ and
\begin{eqnarray}
I\left((Z_k)_{k \in \mathcal{U}_1};(Z_k)_{k \in \mathcal{U}_2} \Big|(Z_k)_{k \in \mathcal{T}}\right) \overset{(\ref{ind_S})}{=} 0 \label{thm3_eq6}
\end{eqnarray}
because $Z_k$ corresponds to all edges incident with node $v_k$ in the hypergraph $H$; conditioning on $(Z_k)_{k \in \mathcal{T}}$ corresponds to the removal of the nodes $\{\left(v_k\right)_{k \in\mathcal{T}}\}$ and the incident edges (keys); all remaining edges $e$ fully lying in $\mathcal{V}_1$ or $\mathcal{V}_2$ leads to that all remaining keys are known either only to users from $\mathcal{U}_1 $ or only to users from $\mathcal{U}_2$. Combining with the fact that the gorupwise keys are independent, we conclude that the above conditional mutual information term is zero.

Next, we show that the above mutual information term must be strictly positive, i.e., the keys must be correlated.
\begin{eqnarray}
    &&  I\left((Z_k)_{k \in \mathcal{U}_1};(Z_k)_{k \in \mathcal{U}_2}|(Z_k)_{k \in \mathcal{T}}\right) \notag\\
   &\overset{(\ref{ind})}{=}& I\left((W_k,Z_k)_{k \in \mathcal{U}_1};(W_k,Z_k)_{k \in \mathcal{U}_2}|(W_k,Z_k)_{k \in \mathcal{T}}\right)\label{thm3_eq1}\\
   &\overset{(\ref{message})}{\geq}& I\left((W_k,X_k)_{k \in \mathcal{U}_1};(W_k,X_k)_{k \in \mathcal{U}_2}|(W_k,Z_k)_{k \in \mathcal{T}}\right)\label{thm3_eq2}\\
   &\geq& I\left(\sum_{k \in \mathcal{U}_1} W_k;(W_k,X_k)_{k \in \mathcal{U}_2}\Bigg|(X_k)_{k \in \mathcal{U}_1},(W_k,Z_k)_{k \in \mathcal{T}}\right)\\
   &=& H\left(\sum_{k \in \mathcal{U}_1} W_k \Bigg|(X_k)_{k \in \mathcal{U}_1},(W_k,Z_k)_{k \in \mathcal{T}}\right)\notag\\
   &&-\underbrace{H\left(\sum_{k \in \mathcal{U}_1}W_k \Bigg|(X_k)_{k \in \mathcal{U}_1},(W_k,X_k)_{k \in \mathcal{U}_2},(W_k,Z_k)_{k \in \mathcal{T}}\right)}_{\overset{(\ref{message})(\ref{corr})}{=}0}\label{thm3_eq3}\\
   &=& H\left(\sum_{k \in \mathcal{U}_1} W_k \Bigg|(W_k,Z_k)_{k \in \mathcal{T}}\right)
   - \underbrace{I\left(\sum_{k \in \mathcal{U}_1} W_k;(X_k)_{k \in \mathcal{U}_1} \Bigg|(W_k,Z_k)_{k \in \mathcal{T}}\right)}_{\overset{(\ref{security})}{\leq}0}\label{thm3_eq4}\\
     &\overset{(\ref{ind})(\ref{h2})}{\geq}& L\label{thm3_eq5}
     \end{eqnarray}
     where in (\ref{thm3_eq1}), we use the independence of the input and key variables and the fact that $\mathcal{U}_1, \mathcal{U}_2, \mathcal{T}$ are disjoint. 
     The second term of (\ref{thm3_eq3}) is zero because $\mathcal{U}_1 \cup \mathcal{U}_2 \cup \mathcal{T} = [K]$ and $\sum_{k \in [K]} W_k$ can be recovered from $(X_k)_{k\in[K]}$. The second term of (\ref{thm3_eq4}) is zero due to the security constraint (\ref{security}). The last step follows from the independence and uniformity of the inputs, and $\mathcal{U}_1 \cap \mathcal{T} = \emptyset$.

Comparing (\ref{thm3_eq5}) with (\ref{thm3_eq6}), we arrive at the contradiction (i.e., $L \leq 0$) and complete the proof that $\mathcal{R}_{gc}^* = \emptyset$ (i.e., secure summation is not feasible).

\subsection{Achievability: $\forall \mathcal{T}, H[\mathcal{V}\backslash\{\left(v_k\right)_{k \in\mathcal{T}}\}]~\mbox{is connected}~\Rightarrow \mathcal{R}_{gc}^* \neq \emptyset$}

We first give an achievable scheme that uses all available groupwise keys (in the same zero-sum manner as in Theorem \ref{thm:groupwise}) and then prove that the scheme is correct and secure.

Suppose the key hypergraph $H = (\mathcal{V}, \mathcal{E})$ contains $K$ nodes $v_1, \cdots, v_K$ and $\left|\overline{\mathcal{G}}\right|$ edges $e_1, \cdots, e_{\left|\overline{\mathcal{G}}\right|}$, where $\overline{\mathcal{G}}$ is the family of the sets of users that share an independent key, i.e., $\overline{\mathcal{G}} = \{\mathcal{G}_1, \cdots, \mathcal{G}_{\left|\overline{\mathcal{G}}\right|}\}$. Then we have $\left|\overline{\mathcal{G}}\right|$ groupwise keys and suppose the key shared by users in $\mathcal{G}_i, \forall i \in [\left|\overline{\mathcal{G}}\right|]$, $S_{\mathcal{G}_i} \in \mathbb{F}_q^{(|\mathcal{G}_i| - 1) \times 1}$ has length\footnote{We will use $|\mathcal{G}_i| - 1$ symbols from $S_{\mathcal{G}_i}$. If we want to make each key has the same length $L_S$, we can set $L_S$ as $\max_i (|\mathcal{G}_i| - 1)$ and zero pad shorter keys.} $|\mathcal{G}_i| - 1$. 

Set $L=L_X = 1$ and the messages as
\begin{eqnarray}
    X_k &=& W_k + \sum_{{\mathcal{G}:k \in \mathcal{G},\mathcal{G}\in \overline{\mathcal{G}} }} {\bf h}_{\mathcal{G}}^{k}S_{\mathcal{G}},~\forall k\in[K] \label{eq:sc1}
\end{eqnarray}
where ${\bf h}_{\mathcal{G}}^{k}$ is a $1 \times (|\mathcal{G}| -1)$
vector set as (suppose $\mathcal{G} = \{u_1,\cdots,u_{|\mathcal{G}|}\} \subset [K]$) 
\begin{eqnarray}
{\bf h}_{\mathcal{G}}^{u_1} &=& [1, 0, \cdots, 0] \notag\\
{\bf h}_{\mathcal{G}}^{u_2} &=& [0, 1, \cdots, 0] \notag\\
&\vdots& \notag\\
{\bf h}_{\mathcal{G}}^{u_{|\mathcal{G}|-1}} &=& [0, 0, \cdots, 1] \notag\\
{\bf h}_{\mathcal{G}}^{u_{|\mathcal{G}|}} &=& [-1, -1, \cdots, -1], \label{eq:hh}
\end{eqnarray}
so
\begin{eqnarray}
    \sum_{k \in \mathcal{G}}{\bf h}_{\mathcal{G}}^k=0. \label{Vsum1}
\end{eqnarray}

Note that for each group $\mathcal{G} \in \overline{\mathcal{G}}$, the sum of all precoded key variables in the messages is zero, so correctness is guaranteed.
\begin{eqnarray}
    \sum_{k \in [K]} X_k &=& \sum_{k \in [K]} W_k + \sum_{k\in[K]} \sum_{{\mathcal{G}:k \in \mathcal{G},\mathcal{G}\in \overline{\mathcal{G}}}} {\bf h}_{\mathcal{G}}^{k}S_{\mathcal{G}}\\ 
    &=& \sum_{k \in [K]}W_k +\sum_{\mathcal{G}:k\in\mathcal{G},\mathcal{G}\in \overline{\mathcal{G}}} \left[\left(\sum_{k \in \mathcal{G}}{\bf h}_{\mathcal{G}}^k\right)S_{\mathcal{G}}\right]
    \overset{(\ref{Vsum1})}{=} \sum_{k \in [K]} W_k.
\end{eqnarray}

Finally, we prove that the scheme (\ref{eq:sc1}) is secure when $H[\mathcal{V}\backslash\{\left(v_k\right)_{k \in\mathcal{T}}\}]$ is connected for any $\mathcal{T}$. 
Consider the security constraint (\ref{security}) for any colluding set $\mathcal{T} \in \overline{\mathcal{T}}$.
\begin{eqnarray}
    &&I\left(\left(W_k\right)_{k\in[K]\backslash\mathcal{T}}; \left(X_k\right)_{k\in[K]\backslash\mathcal{T}}\Bigg|\sum_{k\in [K]\backslash\mathcal{T}} W_k, \left( W_k, Z_k \right)_{k\in\mathcal{T}}\right) \label{eq:ss5}\\
    &=&H\left(\left(X_k\right)_{k\in[K]\backslash\mathcal{T}}\Bigg|\sum_{k\in [K]\backslash\mathcal{T}} W_k, \left( W_k, Z_k \right)_{k\in\mathcal{T}}\right)
    - H\left(\left(X_k\right)_{k\in[K]\backslash\mathcal{T}}\Bigg|(W_k)_{k\in[K]},\left( Z_k \right)_{k\in\mathcal{T}}\right) \notag\\
    &&\\
    &\leq& (K-|\mathcal{T}|-1)- H\left(\left(
    \sum_{{\mathcal{G}:k \in \mathcal{G}, \mathcal{G}\in \overline{\mathcal{G}} }} {\bf h}_{\mathcal{G}}^{k}S_{\mathcal{G}}
    \right)_{k\in[K]\backslash\mathcal{T}}\Bigg|(W_k)_{k\in[K]},\left( Z_k \right)_{k\in\mathcal{T}}\right)\label{lemma:rank_eq2s}\\
    &=& (K-|\mathcal{T}|-1)- H\left(\left(
    \sum_{{\mathcal{G}:k \in \mathcal{G}, \mathcal{G}\in \overline{\mathcal{G}}, \mathcal{G} \cap \mathcal{T} = \emptyset }} {\bf h}_{\mathcal{G}}^{k}S_{\mathcal{G}}
    \right)_{k\in[K]\backslash\mathcal{T}}\right)\label{lemma:rank_eq2ss}\\
    &=& (K-|\mathcal{T}|-1) - (K-|\mathcal{T}|-1) = 0
\end{eqnarray}
where the last step relies on the generic property of the preocoded keys and the connectivity property of $H[\mathcal{V}\backslash\{\left(v_k\right)_{k \in\mathcal{T}}\}]$, and is derived as follows. Suppose there are $M$ groupwise keys that are only known to non-colluding users in $[K]\backslash\mathcal{T}$ and denote the corresponding set of users as $\mathcal{G}_{j_1}, \cdots, \mathcal{G}_{j_M}$, i.e., $\forall m \in [M], \mathcal{G}_{j_m} \in \overline{\mathcal{G}}, \mathcal{G}_{j_m}\cap \mathcal{T} = \emptyset$. Denote $[K]\backslash\mathcal{T} = \{u_1,\cdots,u_{K-|\mathcal{T}|}\}$. Then
\begin{eqnarray}
&& H\left(\left(
    \sum_{{\mathcal{G}:k \in \mathcal{G}, \mathcal{G}\in \overline{\mathcal{G}}, \mathcal{G} \cap \mathcal{T} = \emptyset }} {\bf h}_{\mathcal{G}}^{k}S_{\mathcal{G}}
    \right)_{k\in[K]\backslash\mathcal{T}}\right) \notag\\
    &=& H\left(
    \underbrace{\left[\begin{array}{cccc}
    {\bf h}_{\mathcal{G}_{j_1}}^{u_1} &     {\bf h}_{\mathcal{G}_{j_2}}^{u_1} & \cdots &     {\bf h}_{\mathcal{G}_{j_M}}^{u_1}\\
        {\bf h}_{\mathcal{G}_{j_1}}^{u_2} &     {\bf h}_{\mathcal{G}_{j_2}}^{u_2} & \cdots &     {\bf h}_{\mathcal{G}_{j_M}}^{u_2} \\
        \vdots & \vdots & \ddots & \vdots\\
        {\bf h}_{\mathcal{G}_{j_1}}^{u_{K-|\mathcal{T}|}} &     {\bf h}_{\mathcal{G}_{j_2}}^{u_{K-|\mathcal{T}|}} & \cdots &     {\bf h}_{\mathcal{G}_{j_M}}^{u_{K-|\mathcal{T}|}}
    \end{array}
    \right]}_{=\hat{\bf H}}
    \left[\begin{array}{c}
    S_{\mathcal{G}_{j_1}} \\
    \vdots\\
    S_{\mathcal{G}_{j_M}}
    \end{array}
    \right]
    \right)\\
    &=& \mbox{rank}(\hat{\bf H}) = K - |\mathcal{T}| - 1
\end{eqnarray}
where ${\bf h}_{\mathcal{G}}^u = {\bf 0}$ if $u \notin \mathcal{G}$ and otherwise ${\bf h}_{\mathcal{G}}^u$ is specified in (\ref{eq:hh}). The rank of $\hat{\bf H}$ is $K - |\mathcal{T}| - 1$ because the first $K - |\mathcal{T}| - 1$ rows of $\hat{\bf H}$ are linearly independent (and the last row is the sum of all above rows). Suppose otherwise, i.e., some subsets of the first $K - |\mathcal{T}| - 1$ rows are linearly dependent. Denote the set of the indices of such rows by $\mathcal{G}'$. Then due to the assignment (\ref{eq:hh}), we know that for any column block with subscript $\mathcal{G}_{j_m}$, we must have all $|\mathcal{G}_{j_m}|$ vectors ${\bf h}^u_{\mathcal{G}_{j_m}}, u \in \mathcal{G}_{j_m}$ to produce linearly dependent rows and it must hold that $\forall m \in [M], \mathcal{G}_{j_m} \subset \mathcal{G}'$ or $\mathcal{G}_{j_m} \subset [K]\backslash (\mathcal{T} \cup \mathcal{G}')$. Noting that each $\mathcal{G}_{j_m}$ corresponds to an edge $e_{j_m}$, the above claim means that all edges fully belong to one part of a two-part partition of $H[\mathcal{V}\backslash\{\left(v_k\right)_{k \in\mathcal{T}}\}]$ (whose node sets are given by $(v_{k})_{k \in \mathcal{G}'}$ and $(v_{k})_{k \in [K]\backslash (\mathcal{T} \cup \mathcal{G}')}$), which violates the condition that the induced subgraph $H[\mathcal{V}\backslash\{\left(v_k\right)_{k \in\mathcal{T}}\}]$ is connected. We have thus arrived at the contradiction and have proved that $\mbox{rank}(\hat{\bf H}) = K - |\mathcal{T}| - 1$. 

As the scheme (\ref{eq:sc1}) is both correct and secure and the rate achieved is non-zero, we have proved that $\mathcal{R}_{gc}^* \neq \emptyset$.

\section{Conclusion}
In this work, we have studied an elemental one hop information theoretic model on secure summation. Our main results include the characterization of the capacity region of secure summation with arbitrarily coded keys and symmetric groupwise keys, and the feasibility condition of secure summation with general groupwise keys and colluding user sets. 

\bibliographystyle{IEEEtran}
\bibliography{Thesis}

\end{document}